\documentclass[8.5pt,twoside,twocolumn]{article}
\usepackage[super,sort&compress,comma]{natbib} 
\usepackage{mhchem}
\usepackage{times,mathptmx}
\usepackage{amssymb}
\usepackage{times}
\usepackage{soul}
\usepackage{sectsty}
\usepackage{balance} 

\usepackage{graphicx,color} 
\usepackage{lastpage}
\usepackage[format=plain,justification=raggedright,singlelinecheck=false,font=small,labelfont=bf,labelsep=space]{caption} 
\usepackage{fancyhdr}
\begin{document}

\thispagestyle{plain}
\fancypagestyle{plain}{
\renewcommand{\headrulewidth}{1pt}}
\renewcommand{\thefootnote}{\fnsymbol{footnote}}
\renewcommand\footnoterule{\vspace*{1pt}%
\hrule width 3.4in height 0.4pt \vspace*{5pt}} 
\setcounter{secnumdepth}{5}

\makeatletter 
\def\subsubsection{\@startsection{subsubsection}{3}{10pt}{-1.25ex plus -1ex minus -.1ex}{0ex plus 0ex}{\normalsize\bf}} 
\def\paragraph{\@startsection{paragraph}{4}{10pt}{-1.25ex plus -1ex minus -.1ex}{0ex plus 0ex}{\normalsize\textit}} 
\renewcommand\@biblabel[1]{#1}            
\renewcommand\@makefntext[1]%
{\noindent\makebox[0pt][r]{\@thefnmark\,}#1}
\makeatother 
\renewcommand{\figurename}{\small{Fig.}~}
\sectionfont{\large}
\subsectionfont{\normalsize} 

\fancyfoot{}
\fancyfoot[R]{\footnotesize{\sffamily{1--\pageref{LastPage} ~\textbar  \hspace{2pt}\thepage}}}
\fancyhead{}
\renewcommand{\headrulewidth}{1pt} 
\renewcommand{\footrulewidth}{1pt}
\setlength{\arrayrulewidth}{1pt}
\setlength{\columnsep}{6.5mm}
\setlength\bibsep{1pt}

\twocolumn[
  \begin{@twocolumnfalse}
\noindent\textit{\small{\textbf{Working draft}}}  \newline
\noindent\Large{\textbf{
Collective dynamics in two-dimensional aggregations with competing interactions
}}
\vspace{0.6cm}

\noindent\large{Tamoghna Das \textit{$^{a}$} and M. M. Bandi \textit{$^{b}$}}



\vspace{0.6cm}

\noindent\small{
A generic aggregate forming system in two dimensions (2D) is studied using canonical ensemble constant temperature molecular dynamics simulation. The aggregates form due to the competition between short range attraction and long range repulsion of pair-wise interactions. Choosing the appropriate set of interaction parameters, we focus on characterising the collective dynamics in two specific morphologies, {\it viz.} compact and string-like aggregates. We focus on the temporal evolution of the mobility of an individual particle and the dynamic change in its nearest neighbourhood, measured in terms of the Debye-Waller factor ($\bar{u}_i^2$) and the non-affine parameter ($\chi$), respectively (defined in the text), and their interrelation over several lengths of observation time  $\tau_w$. The distribution for both measures are found to follow the relation: $P(x;\tau_w) \sim \tau_w^{-\gamma}x$ for the measured quantity $x$. The exponent $\gamma$ is equal to $2$ and $1$ respectively, for the compact and string-like morphologies following the ideal fractal dimension of these aggregates. A functional dependence between these two observables is determined from a detailed statistical analysis of their joint and conditional distributions. The results obtained can readily be used and verified by experiments on aggregate forming systems such as globular proteins, nanoparticle self-assembly etc. Further, the insights gained from this study might be useful to understand the evolution of collective dynamics in diverse glass-forming systems.
}
\vspace{0.7cm}
 \end{@twocolumnfalse}
]

\footnotetext{\textit{$^{a}$~ Physics and Applied Mathematics Unit, Indian Statistical Institute, 203 B. T. Road, Kolkata 700108, India. E-mail: dtamoghna@me.com}}
\footnotetext{\textit{$^{b}$~ Collective Interactions Unit, OIST Graduate University, Onna, Okinawa 9040495, Japan. E-mail: bandi@oist.jp}}

{\bf Introduction --}
Strong spatial dependence of local dynamics is generally observed in diverse non-equilibrium systems, \cite{DHrev1,DHrev2,DHrev3} such as molecular and metallic glass formers, polymeric and other network-forming liquids, nanoparticle self-assembly, colloidal dispersion in weakly polar solvents, globular proteins and many more. The origin of such spatio-temporal heterogeneity is commonly attributed to the existence of {\it mobile} and {\it immobile} particles in these systems. Multitudes of computer simulations \cite{crr1,crr2,crr3,crr4} and experiments \cite{dw1,dw2} support the fact that such {\it (im)mobile} particles do {\it not} occur randomly in the system; rather, they exhibit strong correlations over certain finite spatial extent. \cite{string1,string2} This local clustering phenomenon of particles is also dynamic due to the thermal fluctuations present in the system. When a set of closely spaced interacting particles move in unison, the motion of individual particles is expected to be strongly influenced by the motion of their neighbouring particles and vice versa. Over the last few decades, a serious body of research \cite{vsco,DHrev4,dw4,lfs1,lfs2,lem,ml2,lte} has been devoted to identify and characterise such local structures in order to understand the generic features of the collective dynamics and to develop a theoretical framework to describe the same. The questions that remain to be addressed are, how exactly does the motion of an individual particle, involved in collective dynamics, evolve over time? To what extent does the nearest neighbour shell get distorted as the neighbouring particles continue to move collectively under the influence of thermal fluctuations? Most importantly, can we predict the nature and evolution of collective dynamics from the knowledge of individual particle motion? 

This present work attempts to answer these questions by studying the particle trajectories of an aggregate forming system in two dimensions (2D). Particles in this system, upon cooling down at a suitable rate to appropriate temperature, naturally form stable clusters of different shapes and sizes, thereby rendering it a reasonably appropriate model to study the collective particle dynamics in a fluctuating heterogeneous environment. For characterisation of the collective dynamics, we focus on studying the time evolution of two specific quantities. The individual particle motion is quantified by the mobility $\bar{u}_i^2$ defined as the running average of squared displacement of a particle over certain observation time $\tau_w$. The degree of fluctuation of the neighbourhood is measured in terms of non-affine parameter $\chi$ which computes the mismatch between the particle neighbourhoods in the beginning and the end of the same observation time. First, we show that these two observables follow the same scaling form with their observation time confirming the collective nature of local dynamics in this aggregate forming system. Next, we present a phenomenological characterisation of collective dynamics by providing a functional relationship between the response $\chi$ and its predictor $\bar{u}_i^2$ suggested by a master curve obtained from detailed statistical analysis of the distributions of respective quantities.

{\bf Model system and simulation details --}
Formation of aggregates is generally attributed to the competing length scales intrinsic to the particle level interactions. \cite{ci1,ci2} We consider a 2D system of identical particles of unit mass interacting pair-wise via an effective potential $\phi$, an algebraic sum of a short-range attraction, $\phi_{sA}$ and a long-range repulsion, $\phi_{lR}$. The respective functional forms are as follows: $\phi_{sA}=4\epsilon\{(\sigma/r)^{2n}-(\sigma/r)^n\}$ and $\phi_{lR}=(A\xi/r)\exp(-r/\xi)$. The energy, length and time scales, relevant to the simulation are then expressed in units of $\epsilon$, $\sigma$ and $\sqrt{\sigma^2/\epsilon}$, respectively. Following this definition, the repulsion strength $A$ and screening length $\xi$ are expressed in units of $\epsilon$ and $\sigma$, respectively. This specific choice of potentials is a reliable representation of naturally occurring systems, e.g. globular proteins, \cite{gprot1,gprot2} as well as numerous artificially created, yet highly tuneable systems \cite{rev} such as polymer-grafted nanoparticles, \cite{nano} colloids in weakly polar solvent \cite{coll} etc.
\begin{figure}[h!]
\begin{center}
\includegraphics[width=0.95\linewidth]{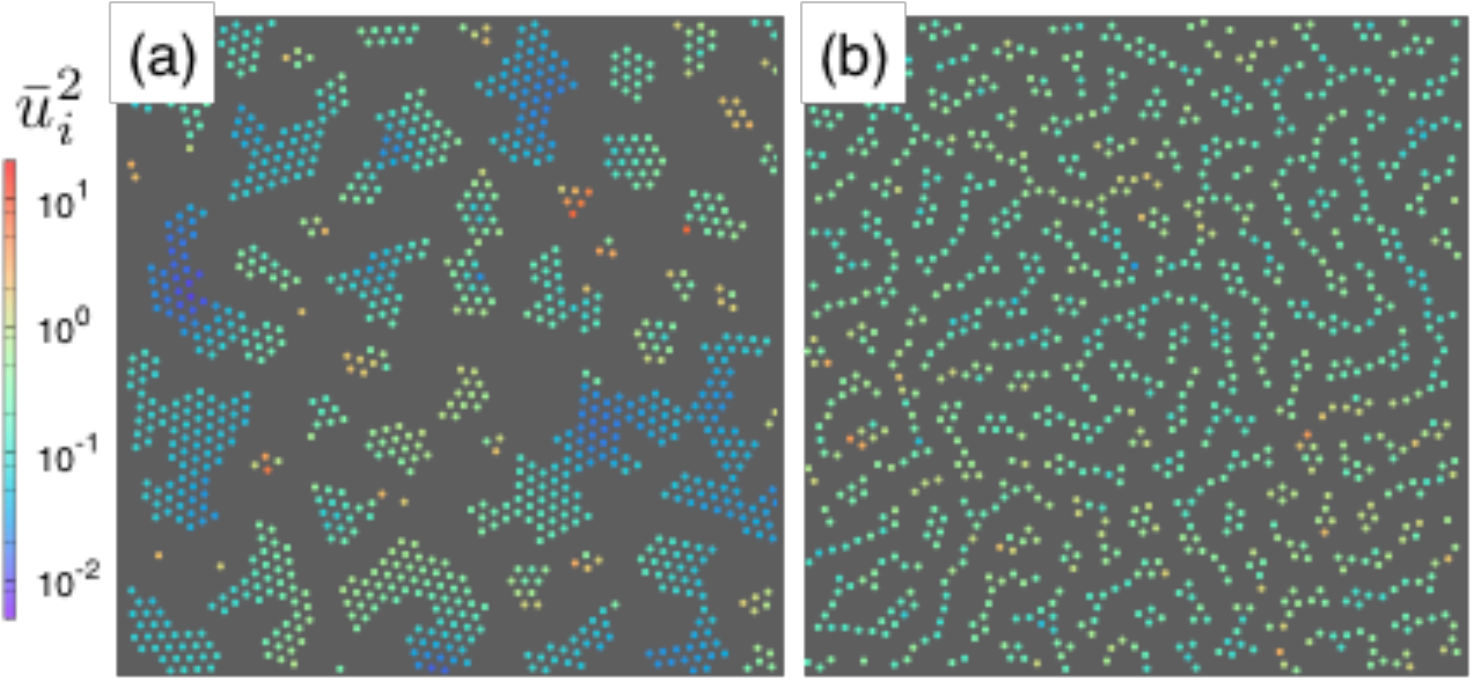}
\end{center}
\caption{(color online) 
Spatial map of particle mobility $\bar{u}_i^2$ at waiting time $\tau_w=50.0$ is shown for different aggregate morphologies, (a) compact cluster ($\xi=0.5$) and (b) string-like clusters ($\xi=0.8$). Only $(1/16)$-th portion of the simulation box is shown here.
}
\label{conf}
\end{figure}

The range of attraction is chosen to be $0.2\sigma$ by fixing $n=18$. For comparison, we state that the attraction ranges up to $2.5\sigma$ for the well-known Lennard-Jones system ($n=6$). By fixing the repulsion strength equal to that of attraction, $A=4\epsilon$, the system's morphology at fixed density and temperature can then be systematically controlled by tuning $\xi$ alone. For the present study, we consider two specific values, $\xi=0.5$ and $0.8$, for which the particles arrange themselves into {\it compact} finite-size crystalline islands with triangular symmetry and as {\it string-like} aggregates, as shown in Fig.\ref{conf}(a) and (b), respectively. We mention that a whole hierarchy of aggregates of intermediate shapes can be obtained by adjusting $\xi$ appropriately. \cite{cism1} The systems are prepared by {\it slowly} cooling an equilibrium liquid of density, $\rho=0.4$ from initial temperature $T_i=1.0k_B\epsilon$ to final temperature $T_f=0.05k_B\epsilon$ with a linear cooling rate of $10^{-4}\epsilon/\tau$. Throughout this canonical {\it constant number-area-temperature} (NAT) simulation, temperature $T$ is maintained using a Langevin thermostat as implemented in LAMMPS \cite{lammps} as already detailed in our previous study. \cite{cism2} The {\it slow} cooling protocol, contrasting the usual rapid quench, is adapted to focus on the role of geometric frustration originating from competing interactions on the anomalous long-time dynamics of the model system. From our previous study, \cite{cism2} we recall that the mean square displacement for both the systems falls out of the ballistic regime after $\tau$ and continues to evolve sub-diffusively at long time after an intermediate relaxation time ranging roughly up to $20\tau$.

{\bf Evolution of particle mobility --} Given the trajectories, we start our analysis by computing the total displacement fluctuation of a particle with respect to its initial position as $\bar{u}^2_i= \int_0^{\tau_w}|{\bf r}_i(t)-{\bf r}_i(0)|^2 dt$. ${\bf r}_i(t)$ is the position of the $i$-th particle at instant $t$ within the measurement period $\tau_w$. Fig.\ref{conf}(a) and (b) show the spatial map of this quantity for compact and string-like aggregates, respectively. Considering this quantity as a reliable measure of {\it mobility} of an individual particle up to $\tau_w$, the distribution of the same $P(\bar{u}^2_i)$ is studied for a set of $\tau_w$ ranging from $\tau$ to $100\tau$.
\begin{figure}[h!]
\begin{center}
\includegraphics[width=0.95\linewidth]{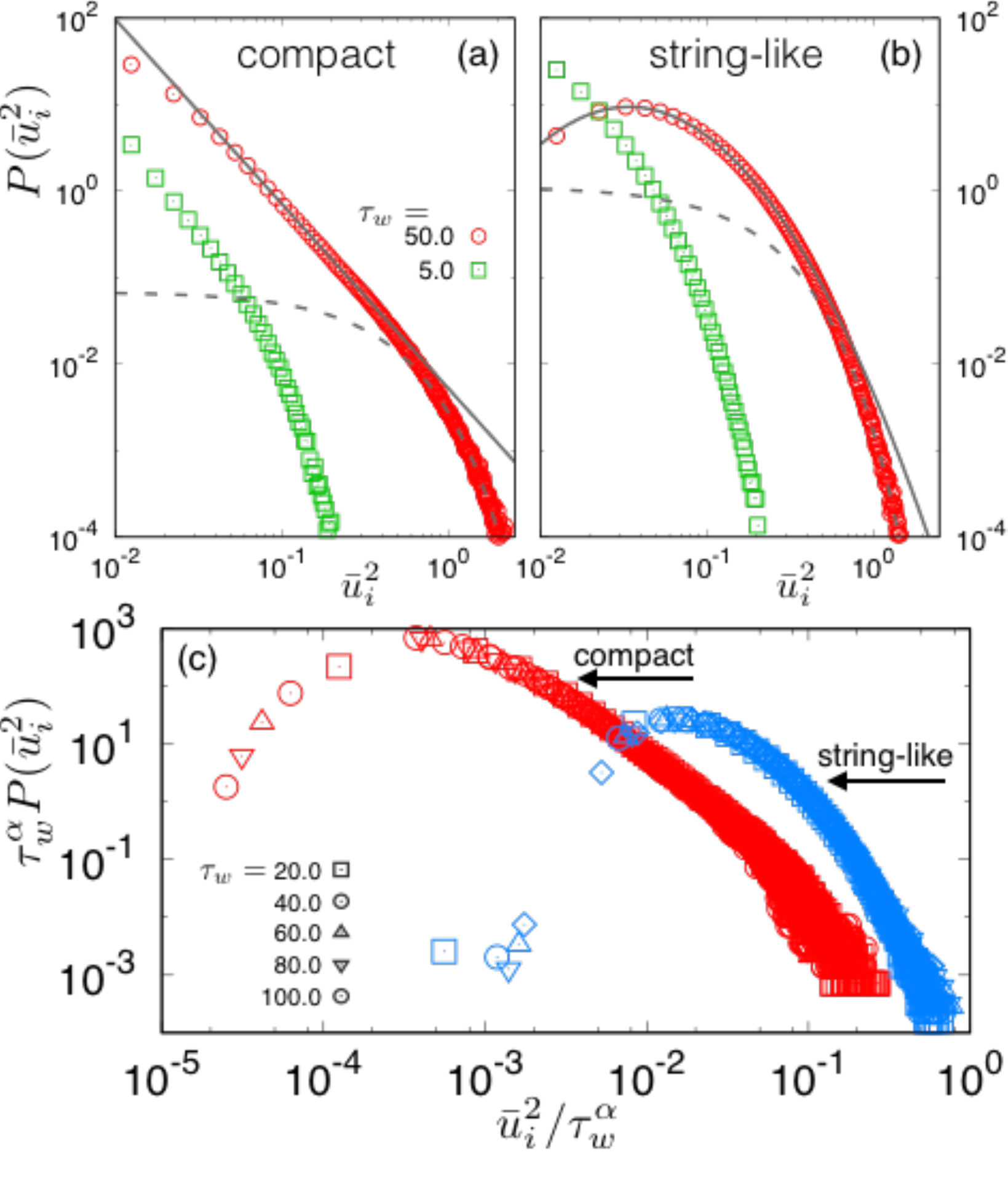}
\end{center}
\caption{(color online) 
(a) and (b) show the distribution $P(\bar{u}_i^2)$ at two different $\tau_w (=5.0, 50.0)$ for the respective cases. The solid line in (a) and (b) are fit to a power-law and a log-normal function, respectively (see text). Both of these distributions develop an exponential tail for large $\bar{u}_i^2 \gtrsim 1.0$. (c) A simple scaling form, $\tau_w^\alpha P(\bar{u}^2_i)\sim\bar{u}^2_i/\tau_w^\alpha$, holds for both compact and string-like morphologies. For compact clusters, the exponent $\alpha$ is unity and $\alpha=0.5$ for string-like aggregates.
}
\label{dw}
\end{figure}
The effect of the shape of the aggregates is evident for both the studied morphologies, $P(\bar{u}^2_i)$ differs from the normal distribution expected for equilibrium liquids. We find that the {\it mobility} of particles forming compact crystalline islands is largely distributed in a scale-free way as plotted in Fig.\ref{dw}(a) and the algebraic portion of $P(\bar{u}^2_i)$ gets extended over larger range of $\bar{u}^2_i$ with increasing $\tau_w$ keeping the exponent unchanged. For string-like aggregates, $P(\bar{u}^2_i)$ is distinctly different and at long time, it closely follows log-normal statistics over almost two decades as shown by the solid line in Fig.\ref{dw}(b). We have confirmed that the system does {\it not} show any overall drift as might be expected from the prominent non-zero peak of the distribution for long $\tau_w$. Due to finite observation time, both of the distributions gets cut off at certain values of $\bar{u}^2_i(\tau_w)$ and decay exponentially as shown by dashed lines in Fig.\ref{dw}(a) and (b).

The systematic change of $P(\bar{u}^2_i)$ with respect to $\tau_w$ can be cast in a simple scaling form: $\tau_w^\alpha P(\bar{u}^2_i)\sim\bar{u}^2_i/\tau_w^\alpha$ which holds for both compact and string-like morphologies. With the exponent $\alpha=1.0$ and $0.5$ for the respective cases, scaled $P(\bar{u}^2_i)$ for all $\tau_w$ collapses to a single master curve as shown in Fig.\ref{dw}(c). Rearranging, we get $P(\bar{u}^2_i;\tau_w)\sim \tau_w^{-\gamma}\bar{u}^2_i$. $\gamma$ is a morphology dependent exponent equal to the ideal fractal dimension $d_f$ for compact ($d_f=2$) and string-like ($d_f=1$) aggregates. This result supports a long term theoretical speculation \cite{trjgeo1, trjgeo2, trjgeo3} of the fractal nature of particle trajectories constituting a fractal object, i.e. aggregates and can be reasoned when {\it mobility} is conceived as the volume swept by the particular trajectory up to the measurement time. However, our study is inadequate to comment further on this issue. Reserving this question for further exploration, we move on to investigate the change in the neighbourhood of $i$-th particle due to the collective motion of its neighbouring particles.

{\bf Evolution of neighbourhood fluctuation --} Unique identification of nearest neighbors in a heterogeneous system is difficult. The traditional way of choosing a cutoff radius equal to the first minimum of the radial distribution function of particles is clearly inadequate to capture the spatial heterogeneity of the system. Purely geometric ways, such as Voronoi analysis \cite{vt0,vt1} or the solid angle nearest neighbor estimation,\cite{solang} have also been reported to produce erroneous results. We employ an adaptive neighbor search algorithm based on relative angular distance \cite{rad} for the present study. This algorithm starts with a list of prospective neighbors, within an arbitrarily large cutoff, arranged in an ascending order of radial distance relative to the central particle. Only those particles are then identified as nearest neighbors which are not angularly blocked by any other particle in the list based on the solid angle criterion. We note that this method is both geometric and non-parametric and thus suitable for spatially heterogeneous situations such as our model system.

\begin{figure}[h!]
\begin{center}
\includegraphics[width=0.85\linewidth]{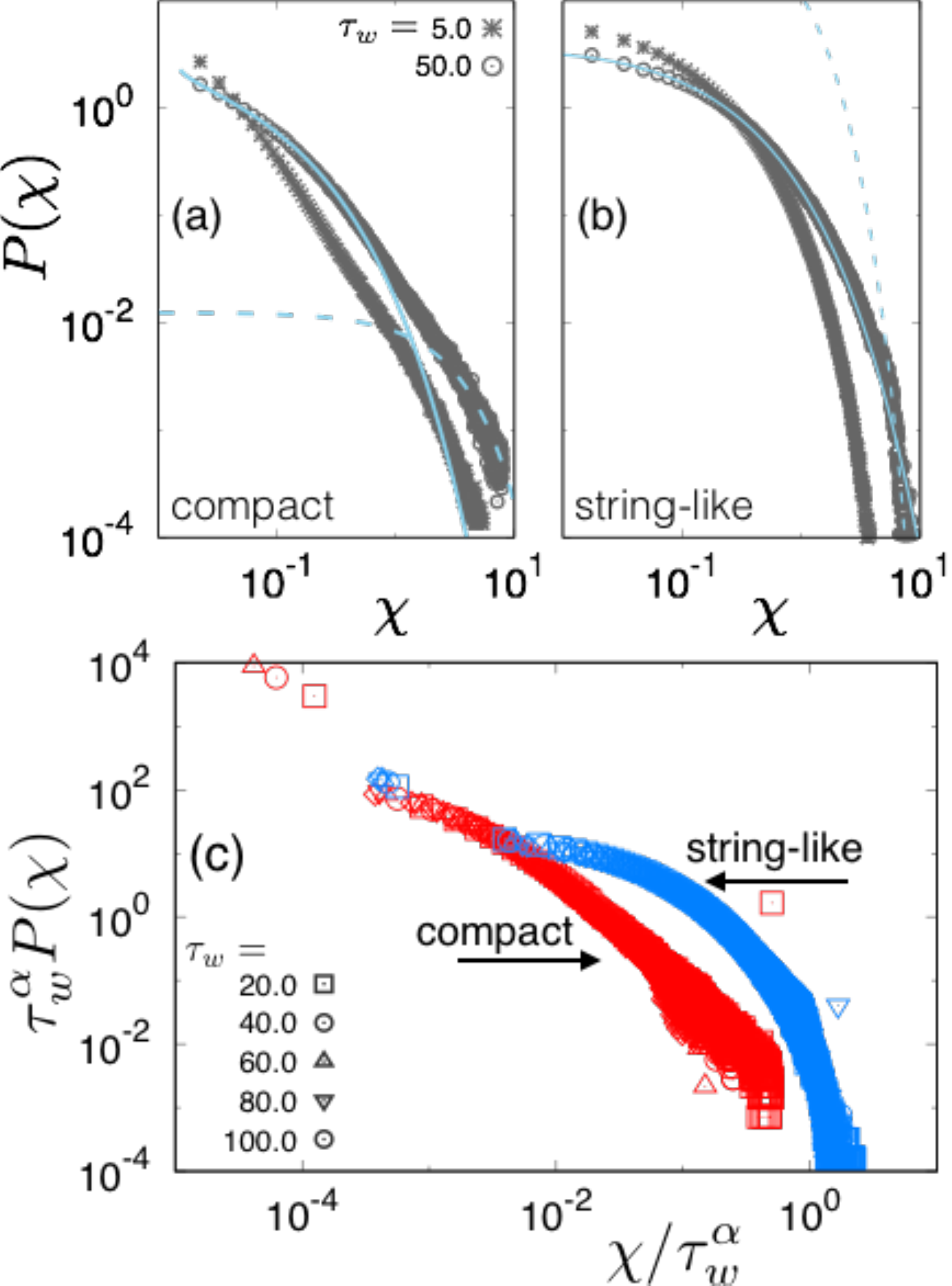}
\end{center}
\caption{(color online)
Probability distribution of non-affine parameter $P(\chi)$ is plotted from (a) compact and (b) string-like aggregates for two different $\tau_w$. The stretched exponential nature of the distribution is shown for both cases with the solid line for $\tau_w=50.0$. The exponential tail, shown by dashed line, is more prominent for the compact case compared to its string-like counterpart. (c) shows the collapse of $P(\chi)$ upon scaling with $\tau_w$ for these two cases separately. Interestingly, the scaling form and exponents are exactly the same as used for $P(\bar{u}^2_i)$ data collapse ({\it see text}).
}
\label{chi}
\end{figure}
Following the identification of first coordination shell around each particle, we compute the non-affine parameter \cite{naffp} defined as $\chi=\sum_j [\boldsymbol{\Delta}_j(\tau_w)-\mathfrak{D}\boldsymbol{\Delta}_j(0)$] where $\boldsymbol{\Delta}_j(={\bf r}_j-{\bf r}_i)$ is the relative distance between the $i$-th particle and its $j$-th neighbor. The matrix $\mathfrak{D}$ attempts to match $\boldsymbol{\Delta}_j$ at any instant to the same at one arbitrary reference time through minimal affine deformation. We note that the restriction of {\it nearest} neighbors was not strictly posed in the original definition. However, by doing so, we eliminate the arbitrariness in the choice of neighbors which might have affected the computation. Further, this restriction returns $\chi$ with a well-defined geometric quantity, namely, {\it procrustean distance} which measures the degree of coincidence between an arbitrary simplex (any nearest neighbor shell, in our case) and a reference simplex (the same shell at initial time). The probability distribution of non-affine parameter $P(\chi)$ computed at the end of two different measurement window, $\tau_w=5.0, 50.0$ is presented in Fig.\ref{chi}(a) and (b) for the compact and string-like aggregates, respectively. While stretched exponential feature, shown by solid line, is observed for both cases, compact aggregates tend to show exponentially distributed large $\chi$ fluctuations (shown by dashed line) compared to string-like aggregates. For comparison, we mention that a clean two-exponential behavior of $P(\chi)$ is observed for model 2D equilibrium liquids. It has been attempted recently to express the non-affine fluctuations as the effect of anharmonic fluctuation arising in a finite temperature system as it explores the potential energy landscape. Following this view, we expect the stretched exponential behavior as an outcome of underlying rugged energy landscape available to our model system. Further study in this direction is underway and will be presented elsewhere.

Interestingly, $P(\chi)$ computed at different $\tau_w$ shows temporal dependence very similar to that of $P(\bar{u}^2_i)$. Using the same scaling relation $P(\chi;\tau_w)\sim \tau_w^{-\gamma}\chi$, excellent data collapse (Fig.\ref{chi}c) can be achieved for both compact and string-like aggregates with $\gamma=2.0$ and $1.0$ for the respective cases. Mentioning both $\bar{u}^2_i$ and $\chi$ have the same dimensionality of $[L^2]$, we recall that $\bar{u}^2_i$ is the total displacement fluctuation of an individual particle over certain observation time; whereas $\chi$ is a geometric distance quantifying the overlap between the nearest neighborhood of the same particle computed at the initial and final instances of the observation period. While the strong interdependence between the {\it mobility} of a particle and the evolution of it's neighbourhood is evident from the existence of a single scaling form with time, it also confirms the collective nature of the local dynamics of our model systems. As it is not straightforward to establish an analytic relation between these two observables, we adopt the standard statistical analysis route to do the same.

\begin{figure}[h!]
\begin{center}
\includegraphics[width=0.95\linewidth]{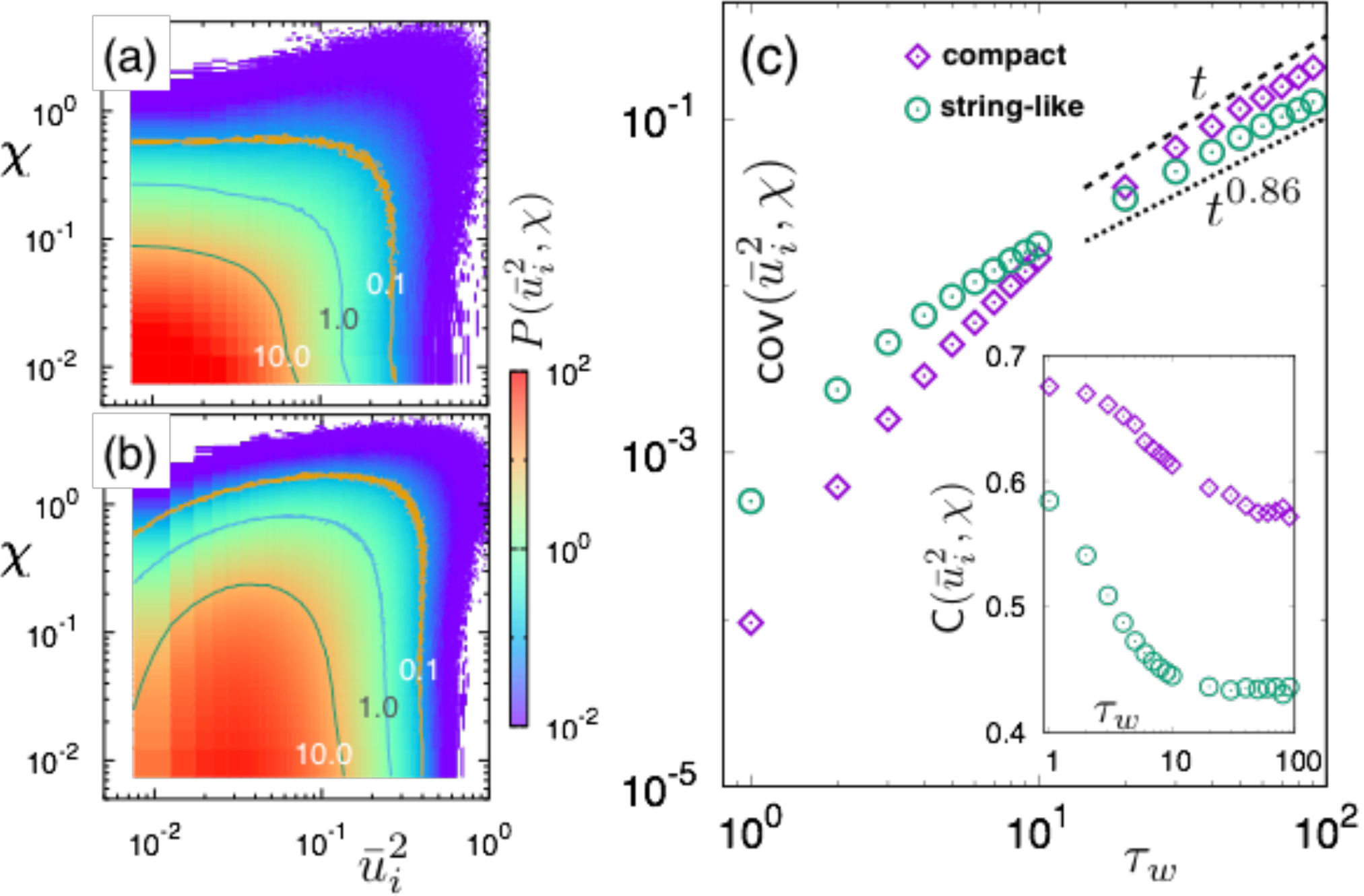}
\end{center}
\caption{(color online)
The joint distribution $P(\bar{u}_i^2,\chi)$ at $\tau_w=10\tau$ for (a) compact and (b) string-like aggregates shows strong interdependence between these two quantities. The iso-lines are drawn as further guide to eyes. (c) The interdependence, quantified by the covariance $\mathsf{cov}(\bar{u}_i^2,\chi)$, grows algebraically at long time when the statistical correlation $\mathsf{C}(\bar{u}_i^2,\chi)$ between these two quantities ({\it Inset}) reaches a steady state at long time.
}
\label{joint}
\end{figure}

{\bf Relation between mobility and neighbourhood fluctuation --} Joint probability distribution is a natural starting point to explore any interrelation between two given random variables. $P(\bar{u}^2_i,\chi)$, joint distribution of $\bar{u}^2_i$ and $\chi$, computed at $\tau_w=10\tau$ for the two sets of studied morphologies, compact and string-like aggregates, are presented in Fig.\ref{joint}(a) and (b) respectively. From this data, it is visually evident that any small change in particle mobility $\bar{u}^2_i$ results into a slight deformation in its nearest neighbor shell, quantified by $\chi$, and vice versa. However, the shape of the distributions suggests that the very nature of interdependence is dictated by aggregates' morphology. To quantify the time evolution of this dependence, we compute the covariance and correlation of the two observables at different observation time $\tau_w$. By denoting the expectation value of a random variable (r.v.) $x$ by $\mathsf{E}(x)$, the covariance, defined as $\mathsf{cov}(\bar{u}^2_i,\chi)=\mathsf{E}(\bar{u}^2_i,\chi)-\mathsf{E}(\bar{u}^2_i)\mathsf{E}(\chi)$, provides information of a particular kind of dependence, namely, the {\em degree of linear relationship} between two observables. The {\em strength of relationship} is provided by the correlation, defined as $\mathsf{C}(\bar{u}_i^2,\chi)=\mathsf{cov}(\bar{u}^2_i,\chi)/\sqrt{\mathsf{Var}(\bar{u}_i^2)\mathsf{Var}(\chi)}$ where the variance of a r.v. $x$ is denoted by $\mathsf{Var}(x)$. We note that a nonzero definite value of either covariance or correlation is indicative of a linear relationship between two r.v.s but a zero value does not exclude the possibility of nonlinear dependence between them. Following Cauchy-Schwarz inequality, covariance is bounded by the standard deviations of individual r.v.s and correlation, by definition, is limited between values $+1$ and $-1$ which are only attained for the cases of strict linear dependence between two r.v.s.

Non-zero values of both the covariance and correlation for all $\tau_w$, as plotted in Fig.\ref{joint}(c) and its {\em Inset}, confirms a strong dependence between $\bar{u}^2_i$ and $\chi$ over all observation times. For small $\tau_w$, $\mathsf{cov}(\bar{u}^2_i,\chi)$ is small and it grows monotonically with increasing $\tau_w$. We consider this as a signature of emergence of collective dynamics within the system. We recall that after time $\tau$, particles in both of our model aggregate forming systems falls out of the ballistic regime as suggested by the appearance of an intermediate plateau  in the mean square displacement data plotted with respect to time. However, over this short timescale, the particles' movement and their fluctuations measured by $\mathsf{Var}(\bar{u}_i^2)$ are expected to be very small due to low temperature and inherent frustration originating from competing interactions. The neighbourhood also remains mostly unchanged as the particles only start to {\em feel} their neighbours after this time leading to small values of $\mathsf{Var}(\chi)$. Consequently, the correlation $\mathsf{C}(\bar{u}_i^2,\chi)$, defined as the covariance over individual variances, shows a large value for $\tau_w\sim\tau$. As time progresses, the collective dynamics sets in leading to increase in both covariance and individual variances of observables which results to the decay of correlation within $\tau<\tau_w<10\tau$. For $\tau>\tau_w$, $\mathsf{C}(\bar{u}_i^2,\chi)$ reaches a steady state value denoting the stabilisation of collective dynamics, namely, {\em caging} for compact aggregates and {\em binding} for non-compact aggregates as revealed by our earlier study. We can then safely consider $\tau_w>10\tau$ as the {\em long time limit} for our purpose. Within this limit, $\mathsf{cov}(\bar{u}^2_i,\chi)$ grows linearly with time for compact clusters and the same shows sub-linear time evolution for non-compact aggregates. The specific reason and implication of such dependence, however, require further detailed investigation which is underway and will be reported elsewhere.

Finally, assuming $\chi$ as a response, we focus to predict a functional dependence of this response on $\bar{u}_i^2$ by computing the conditional expectation $\mathsf{E}(\chi |\bar{u}_i^2)$ as the average value of $\chi$ for a given value of the predictor, $\bar{u}_i^2$. Considering these two observables as discrete r.v.s, formal definition reads as $\mathsf{E}(\chi |\bar{u}_i^2)=\sum_\chi \chi f(\chi |\bar{u}_i^2)$ where the conditional probability density $f(\chi |\bar{u}_i^2)=P(\bar{u}_i^2,\chi)/P(\chi)$ for the non-zero probabilities of the respective quantities. When these two observables are uncorrelated to each other, $\mathsf{E}(\chi |\bar{u}_i^2)$ should return only the value of $\mathsf{E}(\chi)$.
\begin{figure}[h!]
\begin{center}
\includegraphics[width=0.95\linewidth]{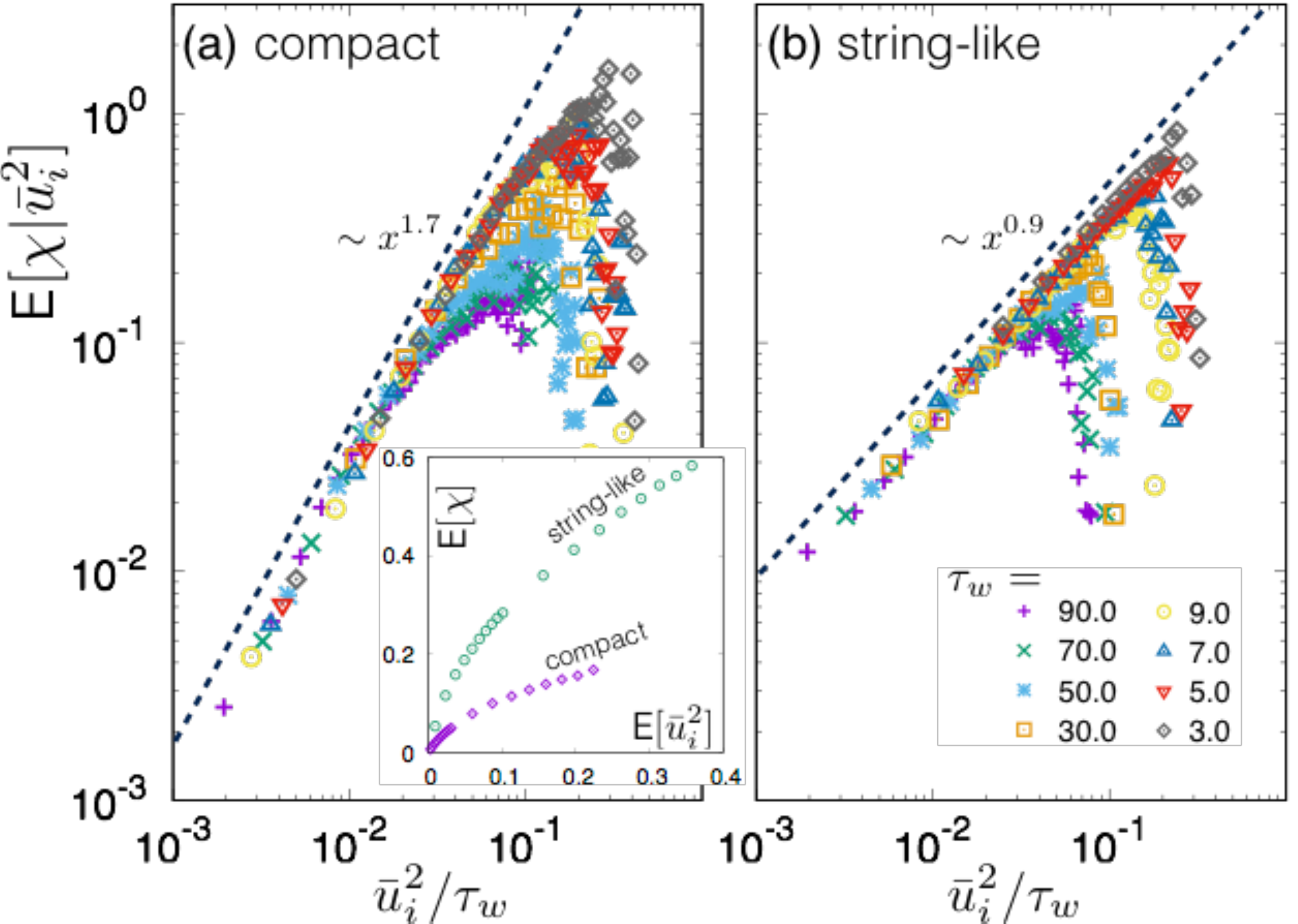}
\end{center}
\caption{(color online)
Scaling of conditional expectation of $\chi$ with respect to $\bar{u}_i^2$ is shown for a range of $\tau_w$ for the case of (a) compact and (b) string-like aggregates. {\it Inset} shows the interdependence between the expectation of these two quantities.
}
\label{conde}
\end{figure}
Now, given the knowledge of $\bar{u}_i^2$, all deterministic functions of $\bar{u}_i^2$ are ideally known since any such function should behave as a constant in terms of the conditional expected value of the same variable. Among all these functions, it can be shown that the conditional expectation is the best predictor of the response as it minimises the mean square error of prediction. This is one basic result of statistics which is valid regardless of the type of the distributions. A non-linear dependence between $\chi$ and $\bar{u}_i^2$ is suggested by their mean values plotted against each other in Fig.\ref{conde}({\em Inset}). The conditional expectation, $\mathsf{E}(\chi |\bar{u}_i^2)$ is found to  show a power-law behaviour with respect to the mobility over unit observation time, $\bar{u}_i^2/\tau_w$ for their small values and drops down after a maximum value which is dependent on the waiting time of observation. This observation prompts us to conclude the following simple functional relation: $\chi \sim (\bar{u}_i^2/\tau_w)^\beta$ with the exponent $\beta$ following the fractal dimension $d_f$ of aggregates very closely. To be specific, $\beta$ values extracted from fitting the data shown in Fig.\ref{conde}(a) and (b), are $1.67$ and $0.88$, respectively, for compact clusters with ideal $d_f=2$ and string-like clusters with ideal $d_f=1$. This functional form might be considered as the best predictor of response $\chi$ for any small change in $\bar{u}_i^2$ at least in regression sense.

{\bf Discussion and conclusion --} 
In summary, we have analysed the simulated particle trajectories of two types of aggregate forming systems, namely, compact and string-like aggregates over several different lengths of observation time. While correlated particle movements are expected in such non-equilibrium systems, this is confirmed by our analysis and the very nature of this correlation is characterised in detail. The mobility $\bar{u}_i^2$ of particles, quantified by the sum of squared displacement over certain observation time, is found to follow a log-normal distribution for string-like aggregates and is scale-free for compact clusters. Irrespective of this morphology dependent difference in the nature of $\bar{u}_i^2$ distribution, time evolution of both distributions follow a common power-law form with observation time with the power-law exponent being close to $2$ and $1$, the ideal fractal dimension of the compact and string-like aggregates, respectively. Structural change of the nearest neighbour shell due the individual particle motion, measured by non-affine parameter $\chi$, is observed to evolve over certain observation time in exactly the same way as $\bar{u}_i^2$, although the $\chi$ distributions are found to follow a stretched exponential form for all observation times and for both compact and string-like aggregates.

This scaling form of the time evolution followed by both $\bar{u}_i^2$ of a single particle and $\chi$, resulting from all its neighbouring particles is a definitive proof of the collective nature of local dynamics of the aggregate forming systems. This is one central result of our study which has not been discussed earlier in literature to our knowledge. The fact that the scaling exponents are same as the fractal dimension of the respective aggregates is remarkable as it suggests an intimate relation between the local structure, specifically geometry, and the local dynamics. A detailed study of the trajectory geometry might be illuminating in this regard and we plan to carry out the same in future. The mobility, $\bar{u}_i^2$ is experimentally measurable \cite{dwexp} and is widely used to study the heterogeneity in local dynamics of a variety of glass-forming systems. \cite{bea1, bea2} In recent years, $\chi$ has been proved to be a useful tool to understand the mechanical response of both crystalline and amorphous media with and without external perturbation.\cite{napre,nasrep,napnas} This study provides a predictive phenomenological relation between these two observables from detailed statistical analysis of their distributions and thus, taking care of all fluctuations of these variables. Such functional relationship, although subject to further scrutiny and verification, should then be of general interest to the soft materials research community.


\end{document}